# Reducing energy consumption of cloud data centers using proper placement of virtual machines

Hamid Reza Naji
Dept. Computer Engineering and Information Technology
Graduate University of Advanced Technology
Kerman, Iran
naji@kgut.ac.ir

Reza Esmaeili
Dept. Computer Engineering and Information Technology
Graduate University of Advanced Technology
Kerman, Iran
r.esmaeili@kgut.ac.ir

*Abstract*— **In today's world, the use of cloud data centers for easy access to data and processing resources is expanding rapidly. Rapid technology growth and increasing number of users make hardware and software architectures upgrade a constant need. The necessary infrastructure to implement this architecture is the use of virtual machines in physical systems. The main issue in this architecture is how to allocate virtual machines to physical machines on the network. In this paper we have proposed a method to use virtualization for minimizing energy consumption and decreasing the cloud resource waste. We have used learning automata as a reinforcement learning model for optimal placement of virtual machines. The simulation results show the proposed method has good performance in reducing energy consumption of servers in cloud data centers.**

## I. Introduction

Virtualization is among the technologies with prominent improvement in systems' efficiency/performance and productivity during the recent decades. Reduction of expenses is generally the main reason to employ virtual technology in organizations [1,2,3,4]. Energy consumption in large-scale cloud data centers is considerably increasing. That's why there is an essential need to improve energy performance/efficiency in such data centers using server consolidation which aims to minimize active physical machines in a data center. The effective methods of placement and migration of virtual machines act as the main key for optimal consolidation [5]. In continue, virtualization and its applications are discussed. The research background is provided. The proposed method and its simulation results are presented. Finally, we can see the conclusion and future works.

## II. Background

Virtualization is the process of management and allocation of virtual resources to various services so that the application programs can utilize the virtual resources created in the platform of real resources [6]. Virtualization technology separates software performance from hardware platforms and segregates the operation from location to present faster network services [7]. One of the most important problems concerning virtualized network functions is the network performance for them. Previous researches show that virtualization can result in too much time delay and consequently, if the network layers are slow, it can cause network power instability [8].

Virtual equipment implemented in data centers, are not managed directly by network operators. Virtual machines can be outsourced to the third parties [9]. In 2014, a novel method of energy consumption management was presented by Horri et al. [10]. The mechanism proposed by this reference is to create compatibility between the amount of energy consumption and the amount of service level agreement violation, i.e.

Jhawar et al. [11], have expanded the process of physical resource management and studied many hypotheses of this process. The mentioned method has a major impact on the determination of the general security requirements in the cloud computing environment [11]. Kliazovich et al. [12] have proposed another form of a data center's energy management mechanism. Patil et al. suggested an efficient mechanism for energy management of a data center's network switches [13]. In 2016, Feller et al. [14], discussed concepts of using cloud environments for big data analysis.

In 2015, Baek et al. [15], suggested a framework, called intelligent frame, based on secure cloud computing for the management of big data information in intelligent networks. In 2017, Garcia et al. [16] offered a solution to energy saving in energy consumption for coordinating and balancing resources in cloud computing. In 2018, Joshi et al. [17] presented an improvement method of energy performance via load optimization in cloud computing.

In reference [18], Genetic optimization algorithm is examined based on the real self-arrangement of virtual machines in a cloud computing data center which is a combination of heterogeneous virtual machines. Genetic algorithm is for optimal allocation of virtual machines in the employed physical resources and the number of virtual machines is dynamic.

## III. THE PROPOSED METHOD

So far, various solutions have been proposed for virtual machines placement and each of them has a specific goal such as symmetric distribution of loads, decreasing energy consumption, etc. In this paper, a new solution to the optimal allocation of virtual machines in the cloud computing system will be presented to provide a low power consumption cloud data center. The proposed method, employs an optimization algorithm to achieve the optimal response. In the following, the proposed method will be presented after explaining the system model. In our model we use a resource optimizer which is explained in bellow:

### A. Resource Optimizer

The Orchestrator needs to be in contact with the Resource Optimizer component to ensure the configuration and proper allocation of virtual machines. The Resource Optimizer component is in charge of selecting service providers' optimal locations in cloud network structure based on the selected services. The main core of Resource Optimizer component is the implementation of algorithms which permit virtual machine allocation to hardware resources available/existing in the network. In the proposed method, the multi-objective cuckoo search optimization algorithm will be implemented in this component. In order to provide this way of optimal placement of virtual machines, the Resource Optimizer component must immediately be able to access the resource condition (cloud resource capacity) and the rate of virtual machine requirements. Then, after determining the way of optimal allocation of resources to virtual machines (by means of the MOCS algorithm in the proposed method), the found optimal solution will be delivered to VM orchestrator.

### B. Optimal Allocation of Virtual Machines

Every virtual machine has characteristics based upon which the allocation operation to the service provider is performed. In the proposed method, three features are considered as the virtual machine's descriptive features i.e. the CPU processing power, the amount of main memory and the storage memory capacity. In case two virtual machines are implemented on one service provider, the amount of service provider's CPU usage will be calculated as the sum of processing power used in the two virtual machines. The total amount of main memory and storage memory used in the service provider will be calculated in the same way. On the other hand, a hundred percent usage of the CPU processing power by virtual machines must be prevented; Since according to the experiments conducted in the previous researches, using the entire CPU processing power in the cloud service providers leads to its instability and performance reduction [19]. In the proposed method, a threshold value will be used to prevent the allocation of physical machines' entire processing power [20].

Besides faster convergence compared with other optimization methods, the cuckoo search algorithm which is an optimization algorithm, doesn't need to determine multiple parameters. The cuckoo search algorithm is an efficient optimization algorithm for solving NP-Hard problems. This algorithm has wide applications in solving the problems related to network and graph domain. The general pattern of this algorithm is to start searching from a set of random solutions and then improve it through different operators. The searching steps via multi-objective cuckoo search optimization algorithm was explained in chapter two. One of the problems of this optimization algorithm is the substitution of improper answers for new random solutions.

As previously stated, in the proposed method, learning automaton units have been employed in order to increase the convergence speed of the optimization algorithm. The learning automata model used in the proposed method employs reward and penalty approach for detection of new optimal placement strategies. During this process, by adjusting the probability of actions based on reward (increasing the selected action probability) and penalty (decreasing the selected action probability) operators, the learning automaton learns which action is optimal and its selection must be made with higher probability in the next cycles. In the proposed method, **N** number of learning automata models will be defined in order to describe the placement strategies of virtual machines in which **N** indicates the number of virtual machines.

Therefore, the pseudocode of the proposed algorithm for the optimal placement of virtual machines based on the combination of the multi-objective cuckoo search optimization algorithm and the learning automata will be as follows:

As it is demonstrated in the above pseudocode, in the proposed algorithm, after determining the initial population randomly, one learning automaton will be made for each virtual machine so that its set of selectable actions is corresponding to the set of physical machines in the problem. The result of this operation will be the formation of **N** learning automata with the selection probability value of $\frac{1}{M}$ for all their actions. Based on the above pseudocode, the process of search and replacement in the proposed algorithm is similar to the multi-objective cuckoo search optimization algorithm; The difference is that, after the placement operation, $\frac{P_a}{2}$ of the worst responses will be replaced by the responses created based on the learning automata's optimal actions. In the same way, $\frac{P_a}{2}$ of the worst responses will be randomly replaced by the multi-objective cuckoo search algorithm.

This operation will be repeated for all virtual machines and the learning automata corresponding to them. Then, the probability of all actions corresponding to the worst response of the current population in learning automata models will be decreased by the penalty operator. This operation is performed for every learning automata model using the relation 1:

$$p_j(k+1) = \begin{cases} (1-b)p_j(k) & j = i, \\ \left(\frac{b}{M-1}\right) + (1-b)p_j(k) & \forall j \neq i. \end{cases} \quad (1)$$

In the above relation, **b** is the penalty factor which is considered equal to 5.0. **M** indicates the total number of learning automata's actions which is equal to the number of physical servers of the problem. Considering this searching mechanism, the response vector structure, the optimization objectives and the problem limitations will be discussed in the following.

### C. Utilization rate of physical resources

The aim of a virtual machine allocation algorithm is the maximum resource utilization of network physical systems. This criterion is calculated as the ratio of allocated physical resources to the total available resources based on relation 2:

$$maximize\ U = \sum_{j=1}^{m}\frac{\sum_{i=1}^{n}(x_{ij}.\alpha R_{p_i})}{T_{p_j}} + \sum_{j=1}^{m}\frac{\sum_{i=1}^{n}(x_{ij}.\beta R_{m_i})}{T_{m_j}} \quad (2)$$

In the above relation, $R_{p_i}$ indicates the amount of CPU required in each virtual machine such as **i** and $T_{P_j}$ indicates the maximum processing power in each server such as **j**. $R_{m_i}$ also shows the amount of main memory required in the virtual machine **i** and $T_{m_j}$ shows the total amount of main memory in the service provider **j**. In the above relation, **α** is the processing resource value required by the client and **β** is the memory resource value required by the client. $x_{ij}$ is the descriptive binary function of the virtual machine **i** allocation to service provider **j**, so that if virtual machine **i** locates in service provider **j**, the function value of $x_{ij}$ will be equal to one and otherwise it will be equal to zero.

### D. Energy consumption

Energy consumption decrease, coinciding with the computing system throughput increase, results in an increase in the system energy performance. So, for the optimal allocation of virtual machines, the energy consumption due to the virtual machine allocation must also be considered as one of the target criteria. This objective can be formulated as relation 3:

$$minimize\ E = \sum_{j=1}^{m}\sum_{i=1}^{n}(x_{ij}.R_{p_i}.P_j) \quad (3)$$

In the above relation, $R_{p_i}$ indicates the amount of CPU required in each virtual machine such as **i** and $x_{ij}$ is the descriptive binary function of the virtual machine **i** allocation to the service provider **j**. $P_j^{idle}$ also shows the constant energy consumption of the physical machine **j** in the idle mode and $P_j$ is the physical machine energy consumption based on the processor utilization rate which is calculated by relation 4:

$$P_j = \begin{cases} (P_j^{busy} - P_j^{idle}) \times U_j^P + P_j^{idle}, & if\ U_j^P > 0 \\ 0, & else \end{cases} \quad (4)$$

In the above relation, $P_j^{busy}$ indicates the average energy consumption in the server **j** in busy modes. $U_j^P$ indicates the CPU utilization rate in the service provider which is calculated as the ratio of the utilized resource amount to the total available resources.

### E. limitations of the proposed algorithm

Limitations are equations or inequalities which are placed next to the objective function and represent restrictions of each of the problem variables. We're facing two restrictions in the problem of network virtual machine allocation through the proposed method. We should consider this limitation that each virtual machine has to be implemented merely in one service provider. Actually one service provider must be allocated to each virtual machine. Therefore, no virtual machines without service provision must be present in the response. On the other hand, one can't divide a virtual machine into two or several segments and implement each segment in one service provider. So it can be concluded that a physical service provider must be allocated to each virtual machine for implementation.

## IV. THE PROPOSED METHOD IMPLEMENTATION

In this section, the results achieved from the research implementation will be explored and the proposed method will be analyzed. The proposed algorithm has been implemented by MATLAB software and the proposed algorithm throughput in the network environment has been examined from total energy consumption aspect.
The results achieved from the simulation of the proposed algorithm is compared to the previous methods. In the following, the results achieved from the simulation and the implementation details is explained.

### A. Simulation

A random database consisting of some hypothetical virtual and physical machines has been used in order to evaluate the proposed method. Every virtual machine has some requirements and has implementation ability in the physical machine so that it can meet the process requirements. It is also supposed that all physical machines have the possibility of parallelism. Therefore, every service provider can host more than one virtual machine. Thus, every input problem in the simulation process has been developed by hypothetical data. That's why, at first, **N** number of physical machines with random processing power have been defined in specific intervals. The central processor power of every server has been described as a random number in the interval 4-16 gigahertz and the amount of main memory has been defined as a random number in the interval 8-32 gigabytes. After defining the physical machines, the number of N<K virtual machines with random requirements and in specific intervals have been developed. These virtual machines have been defined in a way that their total requirements in every problem would be at least 90% of the total servers' processing power. Additionally, the requirements of every virtual machine processor and main memory have been defined less than average amount of the ability of these specifications for physical machines. Thus, in the present problem, at least one physical machine corresponding to the requirements of each virtual machine can be definitely found.

During the simulation operation, results of the proposed method for the problems of virtual machine allocation are investigated from different aspects. That is, by changing the number of virtual machines available in the problem, the allocation operation for these machines will be done through the proposed method. At the end of the simulation, the following criteria have been detected to evaluate the results of the proposed method:

The average of the total energy consumption in service providers: The aim of most algorithms of virtual machine allocation in the cloud, is to reach a status of virtual machine allocation which can reduce energy consumption in cloud servers. Because the high energy consumption in the cloud and the environmental consequences and also the resulting financial burden on the providers of these services are challenges of employing the cloud computing systems. Therefore, considering the energy consumption factor in the processes related to the cloud structure optimization can be very effective in its cost reduction. That's why energy consumption is one of the research criteria in the evaluation of the proposed method. In the following, the results achieved from the proposed method implementation will be explained and its throughput will be compared to and contrasted with the previous algorithms.

*B. Simulation results*

Simulation operation has been performed based on the mentioned parameters and scenarios. During conducting the experiments, the number of virtual machines has been changed from 20 to 100 and the six criteria i.e. energy consumption, resource waste, number of the service providers employed, rate of service provider's CPU utilization, load balancing and processing time have been explored. Moreover, the results achieved from the proposed method evaluation have been compared to and contrasted with GABP algorithm in the reference [21] and PSO algorithm in the reference [22]. The method presented in the reference [21], uses Genetic Algorithm and the method presented in the reference [22], uses Particle Swarm Optimization algorithm in order to solve the problem of virtual machine placement in cloud computing resources. The parameters employed in the proposed algorithm are as follows: population size: 100, maximum number of cycles: 500, $P_a$ probability: 0.25

According to the procedure explained in the previous chapter, the proposed method utilizes the multi-objective cuckoo search optimization algorithm in order to solve the research problem. Instead of considering one objective parameter, a multi-objective optimization algorithm, considers a set of goals in order to detect the optimal response to the problem. In this case, the fitness of answers will be described in the form of a Pareto Front. A Pareto Front diagram shows how far the responses achieved by the optimization algorithm, have approached the existing set of goals in the problem. In Fig 1, a sample Pareto diagram in the 100th iteration of the optimization algorithm employed in the proposed method has been illustrated.

We have named our proposed method as LAMOCS (Learning Automata-based Multi-Objective Cuckoo Search) so in simulation diagrams you can recognize it in compare to other methods.

As is evident from these diagrams, the proposed multi-objective optimization algorithm can always discover a set of answers which can minimize all the three optimization objectives simultaneously. It should be noted that in Fig. 1, power consumption is shown in kilowatts. The second criterion of the proposed method evaluation is the energy consumption in service providers.

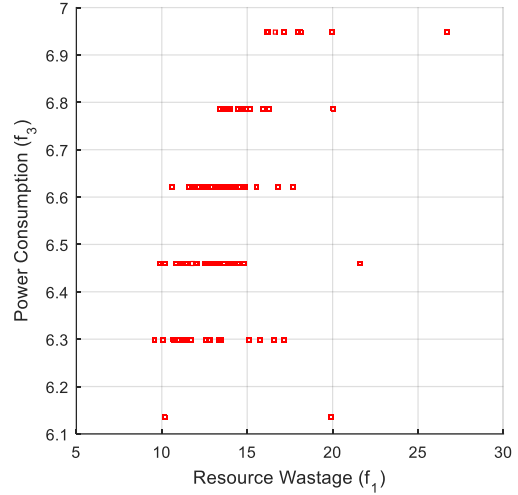

Fig 1. A sample Pareto diagram for power consumption of the proposed method

In Fig 2, the diagram of average energy consumption in service providers for changes in the number of virtual machines is shown.

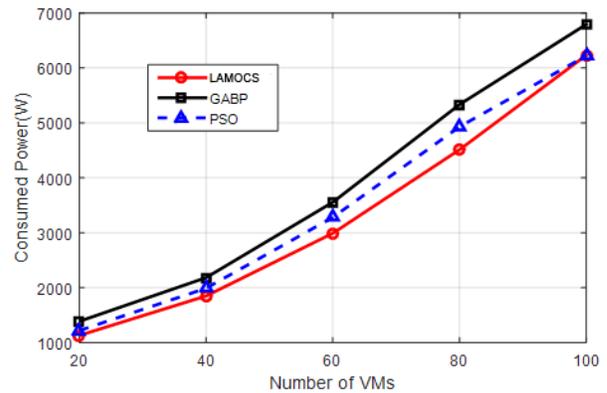

Fig 2. Average of energy consumption in cloud service providers for changes in the number of virtual machines

Based on Fig 2, by increasing the number of virtual machines, the total energy consumption will increase too. Because as specified in the previous experiment, an increase in the number of virtual machines will result in an increase in the number of service providers required for the allocation. Therefore, it's natural that as the number of active service providers in the cloud structure increases, the energy consumption will increase too.

According to the information shown in Fig 2, in all of the allocation situations, through the proposed method, less energy than the compared method will be consumed. Because as mentioned earlier, one of the objectives of the fitness employed in the proposed method, is to perform the allocation operation with the least amount of total energy consumption. The results indicate that the energy consumption in cloud computing systems can be decreased by utilizing the proposed method.

Fig 3 shows the amount of wasted resources in cloud service providers based on the number of virtual machines. Based on Fig 3, an increase in the number of virtual machines results in an increase in the amount of wasted resources in the cloud. Because, as the number of virtual machines increases, the dimensions of the problem increase and accordingly the complexity of the input problem increases. Regarding the fact that, in a problem with **n** number of virtual machines and **m** number of physical machines, the problem dimensions will be equal to $m^n$; Therefore, by the linear increase in the number of virtual machines, (Or: as the number of virtual machines increases linearly, by increasing the number of virtual machines linearly,) the problem dimensions will increase exponentially. This increase in the exponential complexity has caused the diagram of resource waste amount in the cloud, to have an exponential status too.

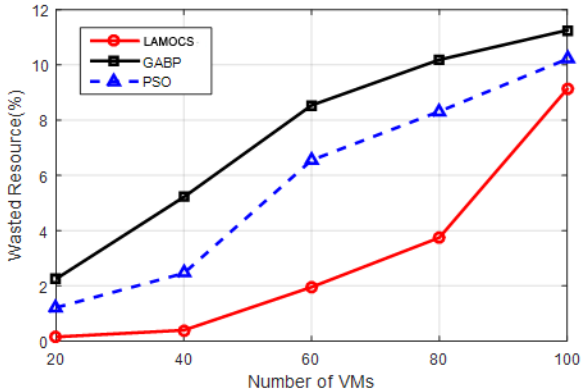

Fig 3. The amount of wasted resources in cloud service providers based on the number of virtual machines.

However, according to Fig 3, in addition to minimum energy consumption, the proposed method has the ability to create responses which result in less amount of resource waste in service providers. The optimization algorithm utilized in the proposed method, through careful investigation of the problem space, chooses a response as the optimal response which has the minimum possible amount of resource waste. In this diagram, the response developed by the proposed method has less resource waste than the other compared methods do. This improvement in the proposed method can be related to better throughput of the multi-objective optimization algorithm via taking advantage of the reinforcement learning strategy of the learning automata.

In Fig 4. The processing time of the virtual machine allocation algorithms is displayed for changes in population size. In this figure the number of virtual machines in equal to 100. Based on Fig 4, on the whole, as population size increases, the processing time of placement algorithms for problem solving increases. Because, by increasing the population size, each optimization cycle will include more computational operations for fitness evaluation of cycle response evolution. Based on the experiments conducted in the present paper, in addition to the performance increase, we have energy consumption reduction in cloud computing systems.

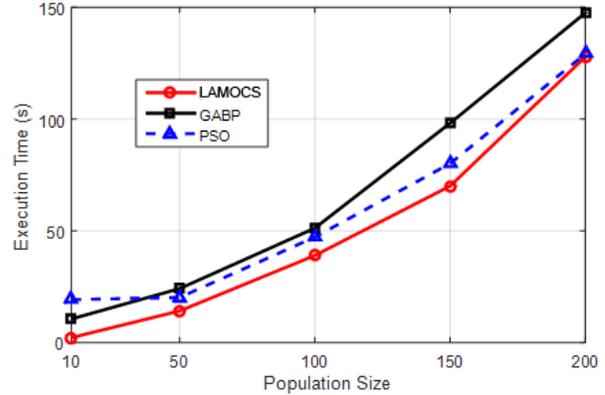

Fig 4. Processing time of the virtual machine allocation algorithms based on population size

## V. Conclusion and Suggestions

### A. Conclusion

In this article, a new approach to the optimal allocation of virtual machines in the cloud data centers has been presented. The method uses energy consumption criteria in order to evaluate the fitness of each answer. Therefore, in this method, the answer with the minimum energy consumption is selected as the final solution. MATLAB software has been utilized to implement the method. Moreover, in the simulation procedure, problems with different sizes (different number of virtual machines) have been used. The results of the experiments conducted, indicate that the answers developed by the method for virtual machine allocation create less energy consumption in cloud resources. In other words, the method selects the allocation cases which lead to energy consumption reduction in the cloud computing system. On the other hand, the method tries to make the most of the cloud service providers' computational capacity. Therefore, the allocations determined by the method will lead to the reduction in the number of active service providers and also the reduction in the cloud wasted resources.

### B. Suggestions

The proposed approach can be utilized to solve similar resource allocation problems in distributed computing systems. Therefore, Examining the application of the method to these problems can be the subject of future studies.